\documentclass[floatfix]{emulateapj}
\usepackage{epsfig,subfigure,amsmath}
\usepackage{color}

\bibpunct[; ]{(}{)}{;}{a}{}{,}

\begin{document}

\title{Bayesian angular power spectrum analysis of interferometric data}
\author{ P.~M.~Sutter$^{1,2,3}$,
         Benjamin~D.~Wandelt$^{2,3,1,4}$ 
         and Siddharth~S.~Malu$^{5}$\\
{~}\\
$^{1}$ Department of Physics, University of Illinois at Urbana-Champaign, Urbana, IL\\
$^{2}$ UPMC Univ Paris 06, UMR7095, Institut d'Astrophysique de Paris, F-75014, Paris, France \\
$^{3}$ CNRS, UMR7095, Institut d'Astrophysique de Paris, F-75014, Paris, France \\
$^{4}$ Department of Astronomy, University of Illinois at Urbana-Champaign, Urbana, IL\\
$^{5}$ Indian Institute of Technology Indore, Khandwa Road, Indore 452 017, India
}
\thanks{Email: psutter2@illinois.edu}

\begin{abstract} 
We present a Bayesian angular power spectrum and signal map inference
engine which can be adapted to interferometric observations of anisotropies in
the cosmic microwave background, 21 cm emission line mapping of galactic
brightness fluctuations, or 21 cm absorption line mapping of neutral hydrogen in
the dark ages. The method uses Gibbs sampling to generate a sampled
representation of the angular power spectrum posterior and the posterior of signal maps
given a set of measured visibilities in the $uv$-plane. We use a mock
interferometric CMB observation to demonstrate the validity of this method in
the flat-sky approximation when adapted to take into account arbitrary coverage
of the $uv$-plane, mode-mode correlations due to observations on a finite
patch, and heteroschedastic visibility errors. The computational requirements
scale as $\mathcal{O}(n_p \log n_p)$ where $n_p$ measures the ratio of the size
of the detector array to the inter-detector spacing, meaning that Gibbs
sampling is a promising technique for meeting the data analysis requirements of
future cosmology missions.  
\end{abstract}


\maketitle

\section{Introduction}

Interferometric techniques have several advantages compared to single-dish
observations: for high resolution experiments they perform the same work of an
optical system but with large potential weight and size reductions, they are
naturally differential and hence well-adapted to measurements of small
anisotropies superimposed on a large isotropic background, their noise
characteristics are highly uncorrelated to a good approximation, and they are
well-suited for observations of statistically isotropic anisotropies, in which
the signal correlations naturally decouple in Fourier
space~\citep{ThompsonA.Richard2001}. Additionally, a revolution in the
development of efficient digital correlator technology, leading to massive
reductions in power requirements~\citep{Parsons2008}, has
brought interferometric approaches back to the list of contenders for
strategies for a future space-based cosmic microwave background (CMB)
mission~\citep{Timbie2009}.

Already, interferometers have made great strides in measuring CMB
anisotropies~\citep{White1999, Halverson2002} and
polarization~\citep{MYERS2006}. The Degree Angular Scale Interferometer
observed anisotropies in the range
$\ell=100-900$~\citep{Halverson2002,Leitch2002} and was the first to discover
CMB polarization anisotropies ~\citep{Kovac2002}.  The Very Small Array 
has observed the CMB at $33$~GHz to an angular scale of
$\ell=1400$~\citep{Grainge2003,Dickinson2004}.  Sensitive to the extremely
small scales of $\ell \sim 4000$ with a resolution of $3-10$~arcminutes, the
Cosmic Background Imager~\citep[CBI;][]{PearsonT.J.2000} has produced a wealth of
information~\citep{Readhead2004,SieversJ.2005,SieversJ.L.2009} and observed the
Sunyeav-Zel'dovich excess at small angular scales~\citep{Mason2003}.  This
excess has also been observed by the SZ Array~\citep{Sharp2010}.

Instruments of the next generation of interferometers are already being
deployed or under active development.
For example,
 the Precision Array for Probing the Epoch
of Reionization~\citep{Parsons2010} and the Murchison Widefield Array
~\citep{Lonsdale2009}
will begin probing the 21 cm regime~\citep{Furlanetto2006, Lidz2008}, while
 the Australian SKA Pathfinder
~\citep{Johnston2008} and the Karoo Array Telescope
~\citep{Booth2009} will study a variety of radio sources.  
The planned Square Kilometer Array
~\citep[SKA;][]{Jarvis2007}
will also be sensitive to 
high-$\ell$ CMB anisotropies~\citep{SubrahmanyanRavi2002}. For any science 
case, the large fields of view, high resolution, and the large number of
antennas pose significant computational challenges to current data analysis
methods.

Current methods for extracting angular power spectrum estimates from observations, such
as maximum likelihood and pseudo-$C_\ell$ (see~\citealt{Tristram2007a} for a
review), have already been applied to interferometric
observations~\citep{Hobson2002, Myers2003} and to more complicated anisotropy
and polarization observing strategies such as drifting and
mosaicking~\citep{Park2003}. However, these strategies are computationally
expensive and scale poorly --- often $\mathcal{O}(n_p^3)$ --- with the data
size $n_p$. Recognizing this problem in observations from bolometer-based imaging instruments, there have
been many efforts to improve the scaling and efficiency of data analysis
algorithms, such as applying massive parallelism~\citep{Cantalupo2009},
adaptive sampling at low-$\ell$~\citep{Benabed2009}, wavelets~\citep{Fay2008},
and Gibbs sampling techniques~\citep{Wandelt2004}.  However, these approaches
have not been extended to interferometers. 

Gibbs sampling is especially promising since it allows a full exploration of
the joint posterior density of the angular power spectrum and signal reconstruction in
$\mathcal{O}(n_p\log n_p)$ operations. This contrasts with the
$\mathcal{O}(n_{p} ^{3})$ scaling for any method that requires evaluations of
the likelihood. Only Hamiltonian sampling has been demonstrated to achieve
similar scaling behavior~\citep{Taylor2008b}.

In the frequentist context, maximum likelihood estimation also scales as
$\mathcal{O}({n_{p}^{3}})$ and is therefore not feasible for the size of current
data sets. Faster estimators have been developed: the optimal quadratic
estimator~\citep{Tegmark1997}, which requires only
$\mathcal{O}({n_{p}^{2}})$ operations, or pseudo-power spectrum estimators
~\citep{Hauser1973,Wandelt2001,Hivon2002,Szapudi2001},
which are designed to approximate this quadratic estimator while costing only
$\mathcal{O}(n_{p}^{1.5})$ operations. 
These tremendous cost savings have rendered the pseudo-power spectrum 
estimators popular choices, although it is unclear how to 
construct controlled approximations of the likelihood starting from such 
an estimator, especially in the regime
of a small number of modes, where a Gaussian approximation 
may not hold~\citep{Elsner2012}. 
Such a likelihood is necessary to perform the Markov Chain Monte Carlos 
with the Bayesian analysis toolkits that have become standard in 
cosmological analysis. In our approach the parameter analysis could be 
performed directly in the context of the power spectrum analysis 
(see the discussion around Equation~(21) in~\citealt{Wandelt2004})
or a controlled, 
convergent likelihood approximation could be built from the Gibbs 
samples themselves~\citep{Chu2005}.
 
Pioneered in the context of CMB observations by~\citet{Jewell2004}
and~\citet{Wandelt2004}, Gibbs sampling has found many applications, such as
Wilkinson Microwave Anisotropy Probe (WMAP) data analysis in temperature~\citep{ODwyer2004, Dickinson2009, Larson2011}
and polarization~\citep{Larson2007, Eriksen2007, Komatsu2011}, and has been
extended to searches in low signal-to-noise regimes~\citep{Jewell2009}. Recently Gibbs sampling  has also been used successfully in combination
with other sampling techniques to reconstruct large-scale structure density
fields and power spectra from galaxy catalogs, modeled as an inhomogeneous
Poisson sample from a normal or log-normal signal~\citep{Kitaura2008,
Jasche2010, Kitaura2011,JascheJens2011}. We will discuss the appropriateness
of Gibbs sampling to observations beyond the CMB in
Section~\ref{sec:conclusions}.

The
technique of Gibbs sampling is applicable even without assuming that the data
are Gaussian. For example,~\citet{Sutton2006} discuss a general approach to
Bayesian image reconstruction from interferometric observations using a fluxon
model. 
Conceptually, the Gibbs sampling approach applied to Gaussian random fields
greatly clarifies the inter-relationship between optimal, minimum-variance
filtering and angular power spectrum estimation. Essentially, Gibbs sampling can be
understood as a non-linear Wiener filter (as discussed
in~\citealt{Wandelt2004}) with the posterior mean map serving as a minimum
variance summary of what can be inferred about the signal without assuming
anything \emph{a priori}. We will return to this subject in
Section~\ref{sec:conclusions}. 

As such, Gibbs sampling neatly resolves an ambiguity in the astronomical data
analysis literature: there are many recipes, but no clear guidance, for
choosing the signal covariance for the Wiener filter~\citep{Rybick1992}. The Bayesian approach implemented through
Gibbs sampling gives a principled and self-consistent way to do angular power spectrum
inference and signal reconstruction at the same time, including full
propagation of the uncertainties. 

In this work we develop such a fully Bayesian analysis as applied to an
observation of the CMB with a simple interferometer, including effects such as
incomplete $uv$-plane coverage, a realistic noise model, and a finite beam
size. Likelihood analysis for realistic interferometer data has been discussed
elsewhere~\citep[e.g.,][]{Myers2003}, but our analysis is novel in the
following ways:
(1) we explore the joint posterior of signal and power spectra; (2) our analysis
   explores the \emph{full posterior shape}; (3) it automatically and fully
takes into account correlations between different points in the $uv$-plane due
to the presence of the primary beam and noise; (4) it does so by generating a
\emph{sampled} representation which makes marginalization trivial (as opposed
to calculating posterior or likelihood slices); (5) we propose a summary of the
signal posterior in terms of the posterior mean, which results in an optimal
signal reconstruction (Wiener filtering) without \emph{a priori} specification
of the signal covariance; and (6) we show how one can build up detailed error
estimates for the reconstruction from the posterior samples.

The analysis reduces to a likelihood analysis for the choice of flat power
spectrum priors, in which case the word \emph{posterior} should be replaced
by \emph{likelihood} throughout the paper.  While our example application of
the technique is simplified---but nonetheless nearly realistic---it
demonstrates the validity of the technique by providing a concrete example and
opens the way for future improvements and applications.
Section~\ref{sec:method} outlines the
method of Gibbs sampling as applied to interferometric observations. Next we
discuss our simulated interferometer observation setup in
Section~\ref{sec:simulations} and present angular power spectrum estimates and signal
reconstructions in Section~\ref{sec:results}. Finally, Section~\ref{sec:conclusions}
summarizes our main findings and discusses the applicability of this 
method beyond the CMB.

\section{Method of Gibbs Sampling}
\label{sec:method}

We model the visibility data $d$ obtained from an interferometric observation as
\begin{equation}
  d = I F A~s + I~n,
  \label{eq:data}
\end{equation} 
where $s$ is a vector containing a discretization of the true sky, $A$ is a
primary beam pattern, $F$ is a Fourier transform operator which converts from
pixel space to the $uv$-plane, $I$ is an interferometer pattern in the
$uv$-plane, and $n$ is a Gaussian realization of the noise. We discuss in more
detail the construction of these elements below in
Section~\ref{sec:simulations}. We discretize the signal $s$, data $d$, and
noise $n$ with $n_p$ elements, using a technique similar to~\citet{Myers2003},
and we assume a flat-sky approximation throughout. 

Gibbs sampling has been extensively discussed elsewhere~\citep[e.g.,][]{Wandelt2004}, so we only briefly
discuss the relevant equations as applied to interferometric observations here.
We begin with some initial guess of the angular power spectrum $C_\ell^0$ and progressively 
iterate samples from the conditional distributions
\begin{eqnarray}
  s^{i+1} \leftarrow P(s | C_\ell^i, m) \\
  C_\ell^{i+1} \leftarrow P(C_\ell | s^{i+1}),
  \label{eq:iterations}
\end{eqnarray}
where $m$ is the least squares estimate of the signal $s$ given the data $d$ 
(i.e., $A^T N^{-1} A m = A^T N^{-1} d$).
The samples $(C_\ell^i, s^i)$ converge to samples from the 
joint distribution $P(C_\ell,s,m) = P(m|s) P(s|C_\ell) P(C_\ell)$ after a 
sufficient number of iterations.

Given a angular power spectrum sample $C_\ell^i$, we generate a new signal sample 
by drawing from a multivariate Gaussian with mean $S^i(S^i+N)^{-1}m$ 
and variance $((S^i)^{-1} + N^{-1})^{-1}$. 
Here $S$ and $N$ are the signal and noise covariance, respectively.
We do this by solving the set of equations
\begin{equation}
\begin{split}
  M~s^{i+1} = & ~A^T F^{-1} I (INI)^{-1} d  \\
              & + F^{-1} S^{-1/2} F~\xi_1 \\
              & + A^T F^{-1} I (INI)^{-1/2} F~\xi_2,
  \label{eq:sky}
\end{split}
\end{equation}
where we define the matrix operator $M$ as
\begin{equation}
M \equiv F^{-1} S^{-1} F + A^T F^{-1} I (INI)^{-1} IFA.
\end{equation}
In the above equations $A^T$ is the beam transpose and $F^{-1}$ is the inverse
Fourier transform. The first term in the right-hand side of the above equation
provides the solution for the Wiener-filtered map, while the second and third
terms of Equation~(\ref{eq:sky}) 
provide random fluctuations with the required variance. The vectors
$\xi_1$ and $\xi_2$ are of length $n_p$ with elements drawn from a standard
normal distribution.

The signal covariance matrix $S$ is diagonal in the $uv$-plane for isotropic
signals, so $S_{\ell,\ell'} = C_\ell \delta_{\ell,\ell'}$, where $\ell = 2 \pi
u$, with $u$ being the radial distance in the $uv$-plane. Here and
throughout we assume the flat-sky approximation that makes this identity
possible. 
By construction, Gibbs sampling explores the exact posterior and therefore 
treats the couplings  introduced by partial sky coverage 
optimally~\citep{Wandelt2004}. The algorithm ``knows'' about the 
couplings since they are contained in the data 
model (Equation~\ref{eq:data}) which underlies the analysis.
However, to a very good approximation the noise covariance matrix $N$ is
diagonal~\citep{White1999}, and thus we will assume this for 
simplicity. We assign to the matrix $N$ entries equal to $N_{i,j} = \sigma_i^2
\delta_{i,j}$, where $\sigma_i$ is the noise variance for the $i$th pixel in
the $uv$-plane. The construction $I (INI)^{-1}$ provides a pseudo-inverse 
 of $N$, so
that any locations in the $uv$-plane with no antenna coverage do not yield
infinities when taking the inverse.  

We solve numerically the above matrix-vector equation using a preconditioned
conjugate-gradient scheme~\citep{PressWilliamH.1986}. The preconditioner
approximates the diagonal components of $M$ and is 
\begin{equation}  
P^{-1} = F^{-1} I (INI)^{-1} I F~(F^{-1} \tilde{A}^2), 
\end{equation} 
where $\tilde{A}$ is the Fourier transform of the beam pattern.

Given the latest signal sample, $s^i$, we generate a new angular power spectrum sample
from Equation~(\ref{eq:iterations}) by computing the variance $\pi_l^2$ in annuli of
constant $\ell$ on the Fourier-transformed signal.  We then use this variance
to draw from the probability density $P(C_\ell | s^i)$, which follows an
inverse Gamma distribution, by creating a vector $p_\ell$ of length $n_\ell$
(assuming a Jeffreys'  ignorance prior) and unit Gaussian random elements.
Here, $n_\ell$ is the number of pixels in the bin $\ell$. The next power
spectrum sample is then simply \begin{equation} C_\ell^{i+1} =
\frac{\pi_\ell}{\left| p_\ell \right|^2}.  \label{eq:clsample} \end{equation}

In the above, we assume $\ell(\ell+1)C_{\ell}$ to be constant across the width
of each annulus in the $uv$-plane. The width of each annulus can be set as
desired. For the example that we show in this paper we chose the width to be $8
\pi/L$, where $L$ is the longest baseline of the interferometer. This is four
times the $uv$-space resolution. This choice limits
correlations between angular power spectrum bins which develop as a consequence of
partial sky coverage. All $\ell$-bins have uniform width except for the first,
which we restrict to cover only the central zone where we enforce $C_0^i = 0$, 
since our analysis cannot constrain the DC mode. We wish to capture 
as much power spectrum information as possible, 
so we correspondingly widen the width of the second bin to close this gap.

To determine convergence so that our iterative samples from the conditional
distributions (Equation~\ref{eq:iterations}) are indeed samples from the joint
distribution, we employ multiple chains with different random number seeds. Our
convergence criterion is the Gelman-Rubin (G-R) statistic, which compares the
variance among chains to the variance within each chain. 
The G-R statistic asymptotes to unity, so convergence is said to
be achieved when this statistic is less than a given tolerance for each
$\ell$-bin~\citep{Gelman1992}.

\section{Simulations}
\label{sec:simulations}

We begin with a realization of the CMB sky on a square patch of side
$20$~deg at 30 GHz 
based on a CAMB-produced angular power spectrum~\citep{Lewis2000} with
cosmological parameters consistent with WMAP seven-year results~\citep{Larson2011,
Komatsu2011} of $\Omega_M=0.27$, $\Omega_\Lambda=0.73$, $\Omega_b=0.045$, and
$H_0=70 {\rm km}~{\rm s}^{-1}~{\rm Mpc}^{-1}$. We discretize this signal map
and corresponding $uv$-plane with $256$ pixels per side, which gives a spatial
resolution of $4.7$~arcmin and a $uv$-plane resolution 
of $2.86~\lambda$. While
this size of patch violates the strict flat-sky approximation, it allows us to
explore the validity of our technique at higher resolutions and probe the range
of scales accessible to realistic interferometers, such as the
CBI~\citep{Mason2003}.

We model the primary beam pattern $A$ as a Gaussian with peak value of unity
and standard deviation $1.5$ deg. 
With these parameters the beam decreases to a value
of $10^{-3}$ halfway to the edge of the box. This allows us to include all
Fourier modes up to the Nyquist frequency in our analysis and ensures that the
periodic boundary conditions inherent in the Fourier transform do not cause
unwanted edge-effects.

\begin{figure} \centering 
  \includegraphics[width=\columnwidth]{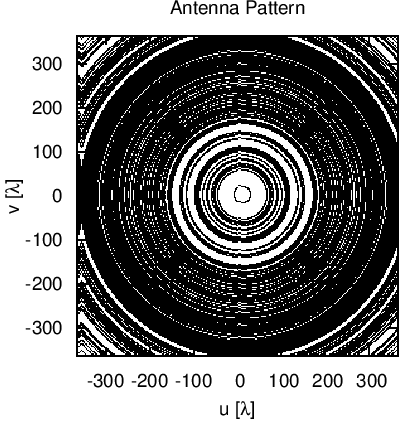}
  
  \caption{Coverage pattern (black) in the $uv$-plane. 
           This pattern results from selecting all baseline pairs from
           20 randomly-placed antennas pointed at a celestial pole. 
           The pattern is then
           rotated uniformly on the same patch of sky for $12$ hr.}
\label{fig:antenna}
\end{figure}

\begin{figure}
  \centering 
  \includegraphics[width=\columnwidth]{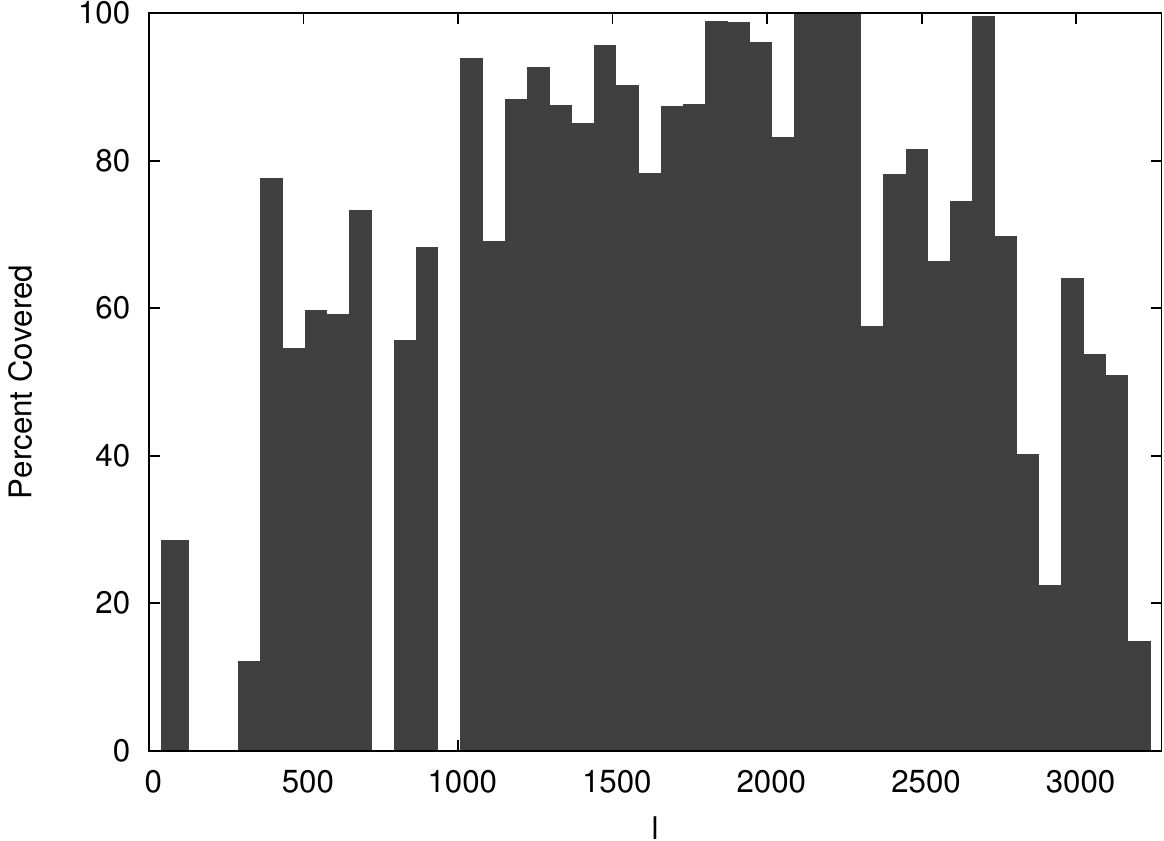}
  
  \caption{Antenna coverage for each $\ell$-bin. 
           Shown is the percentage of cells 
           within the annulus defined by the radius $\ell$
           intersected by any antenna during the $12$ hr observation period.}
\label{fig:skycoverage}
\end{figure}

We assemble the interferometer array in a simple way by randomly placing 
20 antennas and selecting all baseline pairs within the $uv$-plane.
We then allow the assembly to rotate uniformly for
12 hr while observing the same sky patch at a celestial pole. 
We construct the interferometer
pattern $I$ by placing a value of one wherever a baseline length intersects a pixel
during its rotation and zeros elsewhere. We show the resulting $uv$-plane
coverage in Figure~\ref{fig:antenna}.  This configuration covers roughly $70
\%$ of the $uv$-plane, although the coverage varies significantly for each
$\ell$-bin, as Figure~\ref{fig:skycoverage} shows. Some bins, especially at
very low and very high $\ell$, have zero coverage due to the lack of baselines
at that distance. However, even if these bins had adequate coverage, we expect
statistics here to be relatively poor due to the reduced number of modes in
these regions. Most bins have at least $60 \%$ coverage and several bins have
complete coverage.

\begin{figure*}
  \centering 
  \subfigure[Input Sky ($s$)]{
    \includegraphics[width=0.31\textwidth]{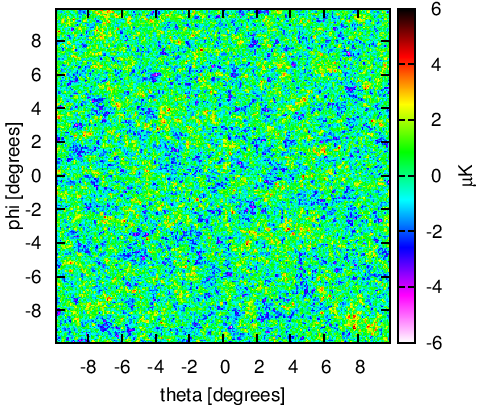}
  } 
  \subfigure[Beam Application ($A~s$)] {   
    \includegraphics[width=0.31\textwidth]{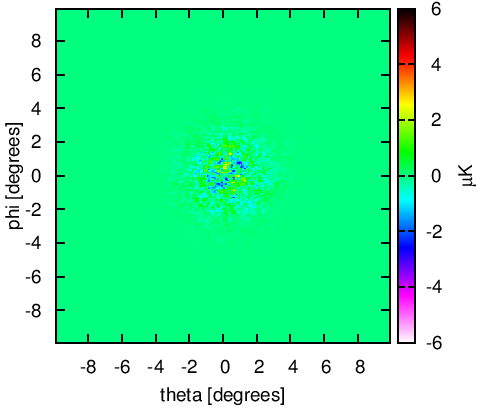}
  }
  \\
  \subfigure[Fourier Transform ($F A~s$)]{
    \includegraphics[width=0.31\textwidth]{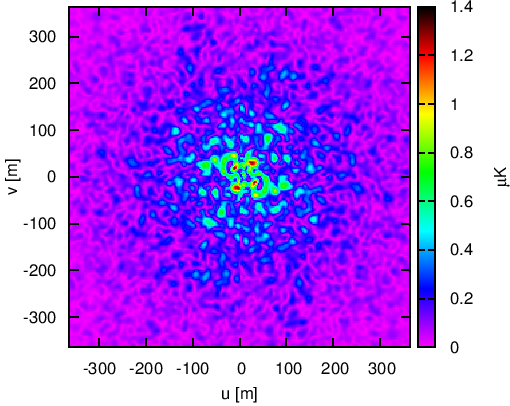}
  }
  \subfigure[Interferometer Application ($I F A~s$)]{
    \includegraphics[width=0.31\textwidth]{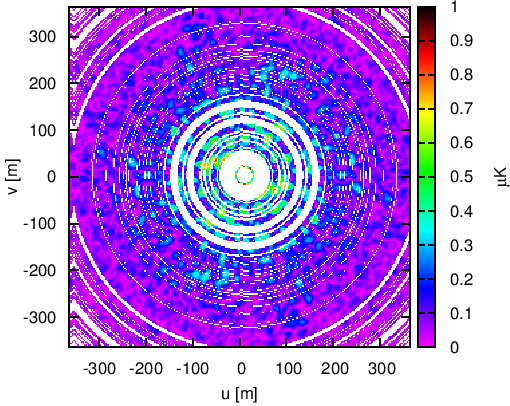}
  }
  \subfigure[Data ($I F A~s + I~n$)]{
    \includegraphics[width=0.31\textwidth]{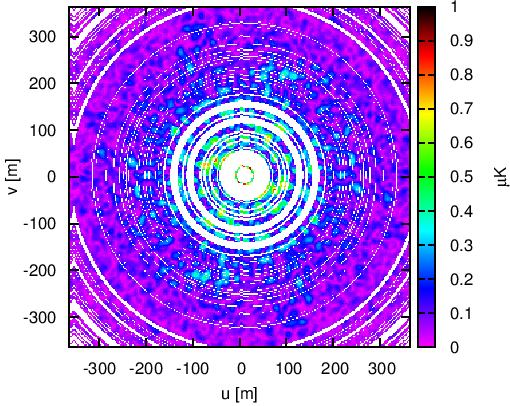}
  }
  \caption{Observation-making process. Shown are
           (a) the input sky signal $s$,
           (b) the application of a primary beam, which we model
              as a Gaussian with standard deviation $1.5$~deg,
           (c) Fourier transformation onto the $uv$-plane,
           (d) application of $20$ randomly placed antennas rotated uniformly
              for 12 hr, and
           (e) the addition of the noise.
           Note that all images in the $uv$-plane are shown as magnitudes.}
\label{fig:data}
\end{figure*}

We determine the noise per pixel by summing the integration time spent in that
pixel by all baselines. We do not adopt a noise model for a 
particular instrument; rather, we set the noise variance 
to be $\sigma_i^{2} \propto
1/t_{{\rm obs},i}$. We then set an overall signal-to-noise ratio of $10$
by multiplying all noise variances by a constant value to maintain
$|IFA~s|/|In|=10$.  This provides a scaling of the noise that would 
normally be caused by instrument effects such as the surface area of the 
detectors and the system temperature in a realistic observation.
We use a Gaussian realization of this noise to create the
data in Equation~(\ref{eq:data}).  We show the process going from input signal to
data $d$ in Figure~\ref{fig:data}. 

\subsection{Computational Considerations and Correlation between Samples}
\label{subsec:comp}
The computational complexity of the Gibbs sampling algorithm as applied to
single-dish experiments has been discussed in detail in previous
works~\citep[e.g.,][]{Wandelt2004} where it was shown that the scaling is
dominated by spherical harmonic transforms, which scale as
$\mathcal{O}(\ell^{3})$.  Since we are dealing with small patches on the sky,
assume that all horns have identical beams, and discretize the $uv$-plane, the
analysis time is dominated by two-dimensional fast Fourier transforms, 
which scale as
$\mathcal{O}(\ell^{2}\log{\ell})$.

The scaling estimate concerns
the time to compute individual samples.  The total time required to perform a
full analysis also depends on the number of samples required for a satisfactory
exploration of the Bayesian posterior density.  For single-dish experiments the
correlation length for individual $C_{\ell}$ bins can become large for high
$\ell$ where many modes with individually low signal-to-noise combine together.
This can lead to very long convergence times and limited scalability, which
requires special tricks for the analysis of low signal-to-noise regimes as
described in~\citet{Jewell2009}. As we will see, however, compact antenna 
arrays result in an approximately uniform coverage of the $uv$-plane. 
The correlation length is a function of coverage and therefore we obtain 
approximately uniform signal-to-noise as a function of $\ell$ which 
means the correlation function will be a weak function of $\ell$. 

\begin{figure}
  \centering 
  {\includegraphics[width=\columnwidth]{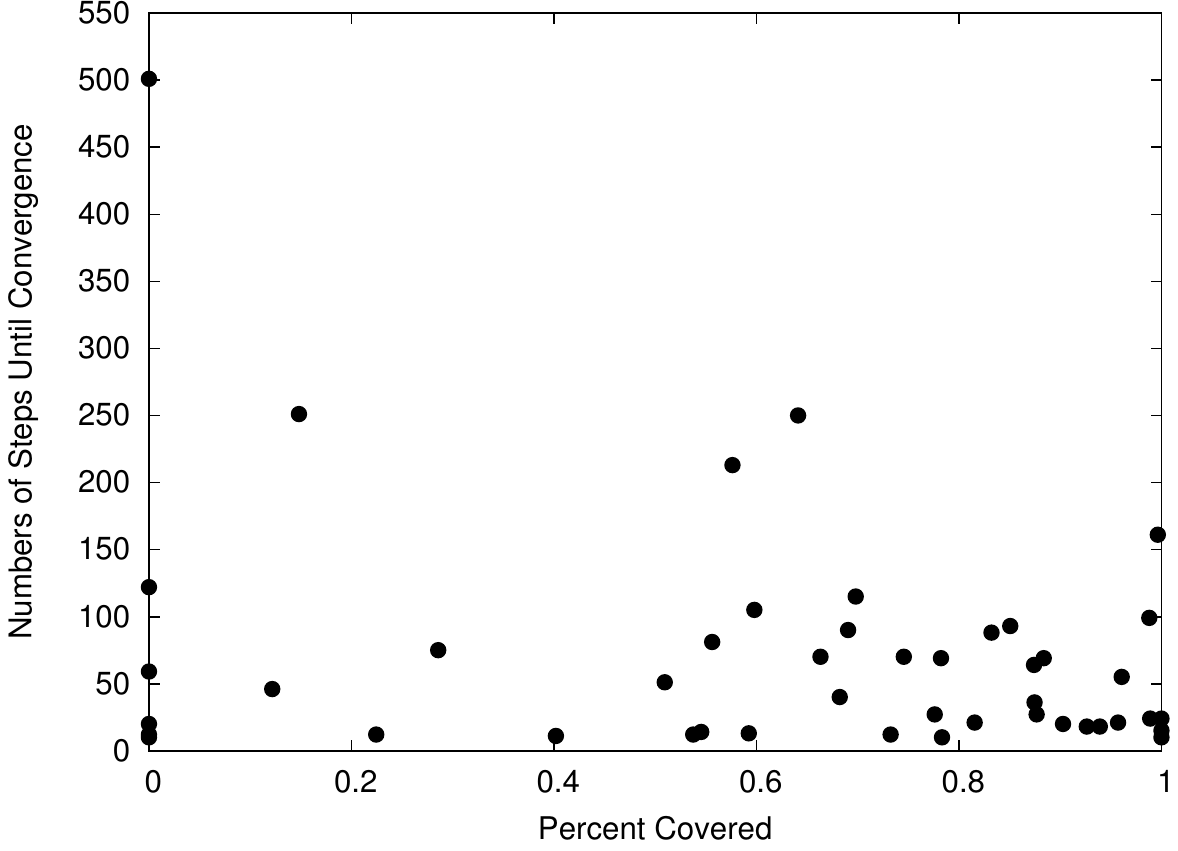}}
  \caption{Number of steps after burn-in required 
           to satisfy the G-R convergence criterion 
           versus the sky coverage percentage for each $\ell$-bin.} 
\label{fig:stepstoconvergence}
\end{figure}

We ran four independent Gibbs sampler chains for $2000$ iterations each.  We
discarded the first $1000$ iterations for a ``burn-in'' phase.  After this
number of iterations, the G-R statistic reached less than $1.1$ for all bins.
This analysis took roughly 75 hr on 4 cores of an Intel dual six-core X5650
Westmere 2.66GHz machine and consumed only 10 MB of memory.  We show in
Figure~\ref{fig:stepstoconvergence} the number of steps required to satisfy our
G-R convergence criterion versus the sky coverage percentage for each
$\ell$-bin.  Shown are the number of steps \emph{after} the burn-in phase.  

We
see little correlation between $uv$-plane coverage and time to convergence.
As we will see in the power spectrum results below, reduced coverage 
does lead to a variable correlation length 
(larger for small power and shorter for large power). 
This means that some low-coverage bins dominate our
convergence rate and hence overall performance. This is a consequence 
of the ratio of the size of the Gibbs move to the full 
width of the posterior~\citep{Erickson2004}.

Large correlation lengths, which lead to longer convergence times, do not have
a large impact on the feasibility of the example analysis we present here;
however, they would be more important in the limit where detector noise is
large compared to the observed signal (e.g., if the goal was to place upper
limits on an as-yet unobserved signal, as is currently the case with
inflationary B-mode missions). This issue will therefore have to be revisited
for polarization data. In any case, if long correlation lengths in the Markov
chain were to cause poor performance, we could follow the general prescription
of~\citet{Jewell2009} by incorporating a step with a 
Metropolis-Hastings sampler 
and deterministic rescaling of the sky signal.

\section{Results}
\label{sec:results}

\subsection{Power Spectrum}

\begin{figure}
  \centering 
  {\includegraphics[width=\columnwidth]{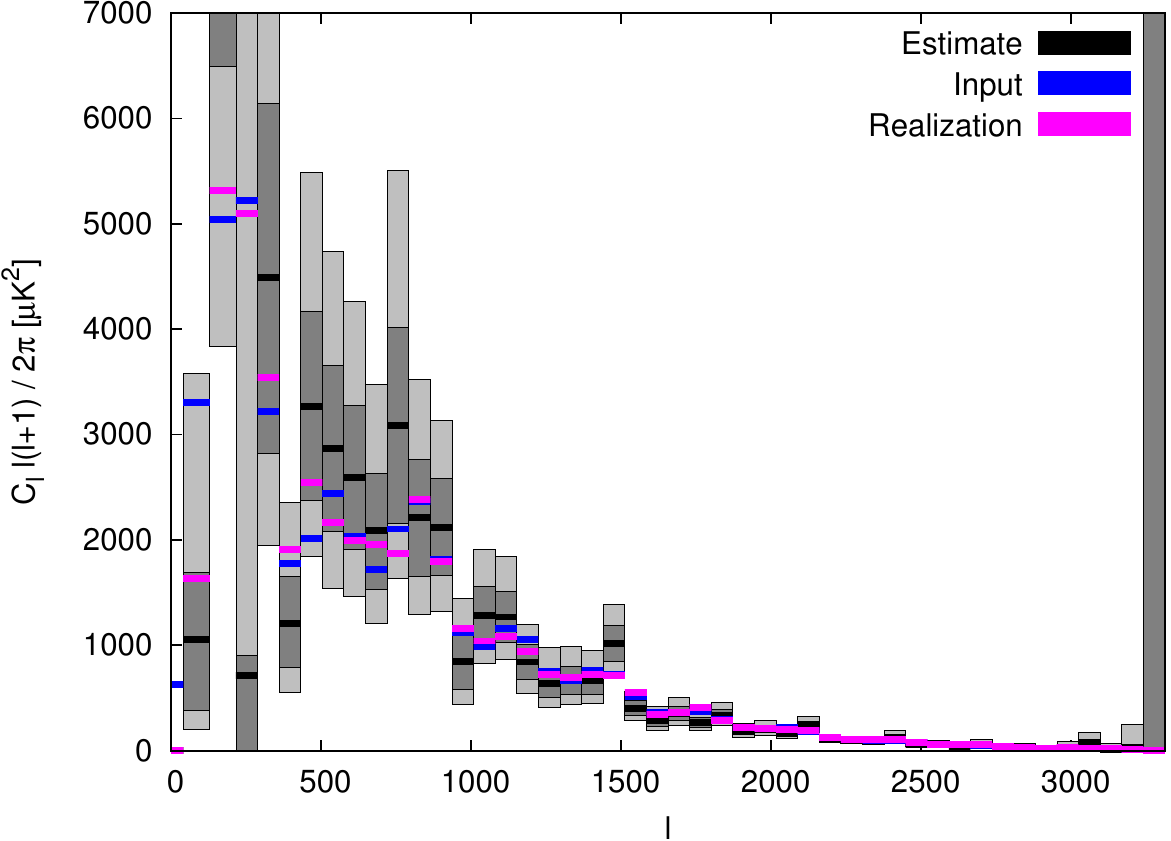}}
  \caption{Mean posterior angular power spectrum for each $\ell$-bin (black) 
             with $1\sigma$ (dark gray) and $2\sigma$ (light gray) 
             uncertainties. The binned input CMB angular power spectrum is shown 
             in blue 
             and the angular power spectrum of our specific 
             signal realization is shown in pink.}
\label{fig:powerspec}
\end{figure}

In Figure~\ref{fig:powerspec} we show the mean posterior angular power spectrum of the
of the four independent chains after reaching convergence. We also show the
uncertainty associated with each $\ell$-bin and the corresponding power of our
input signal realization. The size of the uncertainties in each bin are
consistent with varying coverage in the $uv$-plane; i.e., those bins with
little to no coverage have correspondingly large error bars, while those bins
with complete coverage have the tightest constraints. All of our estimates fall
within $2\sigma$ of the expected value, and most within $1\sigma$, as expected
with this number of bins.  

\begin{figure*} \centering 
  {\includegraphics[width=0.31\textwidth]{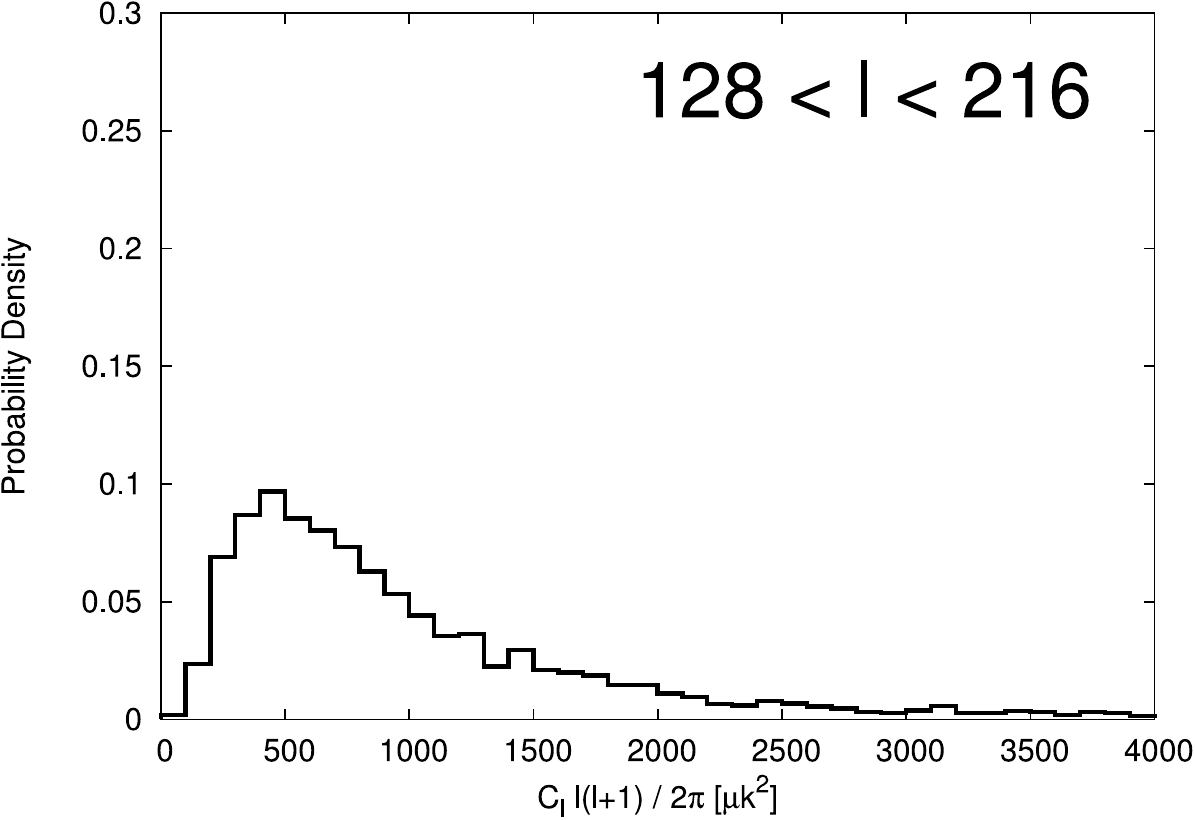}}
  {\includegraphics[width=0.31\textwidth]{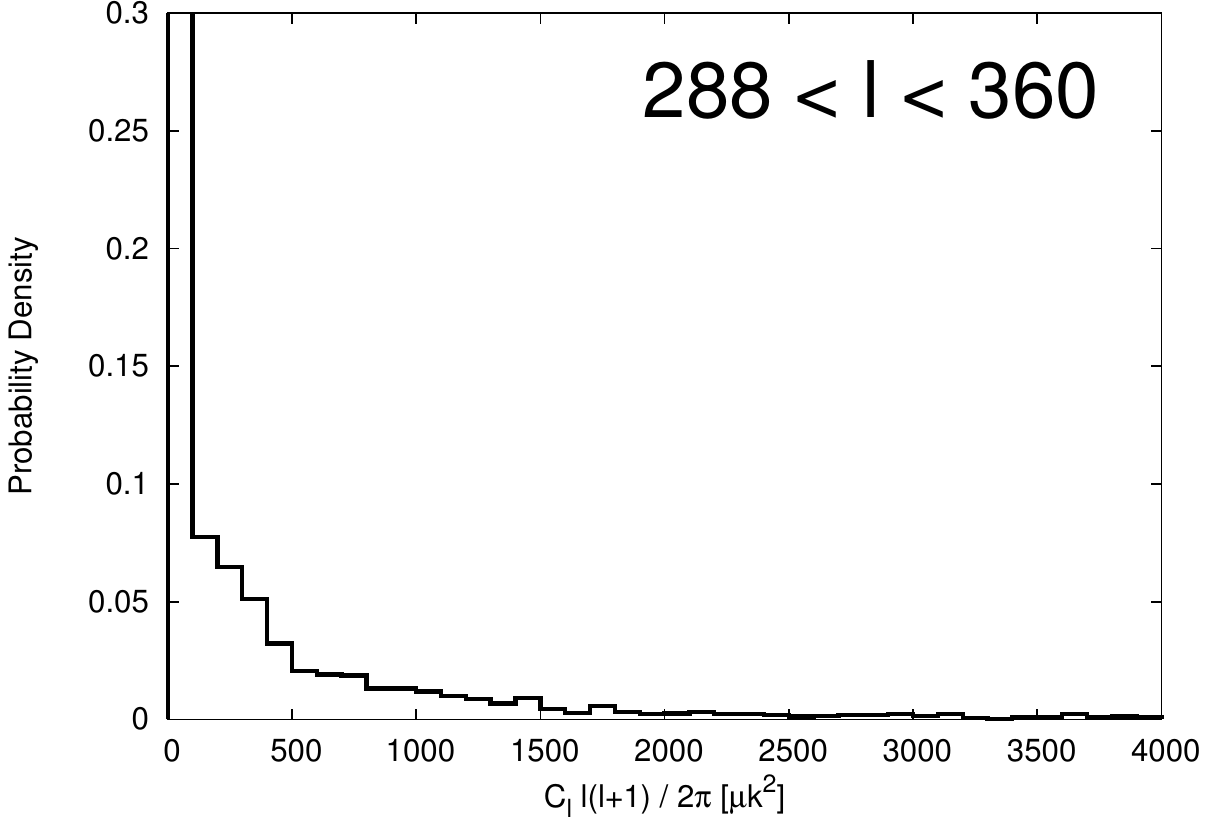}} 
  {\includegraphics[width=0.31\textwidth]{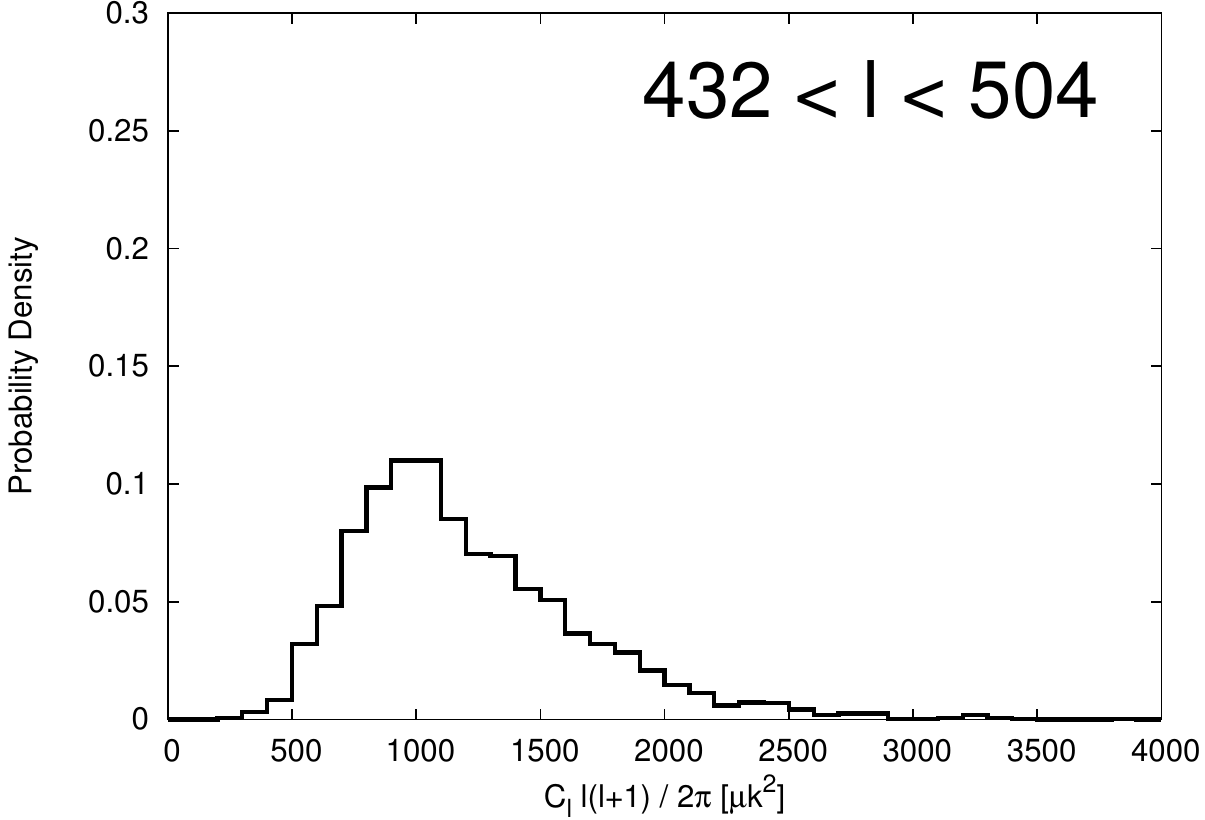}}
   {\includegraphics[width=0.31\textwidth]{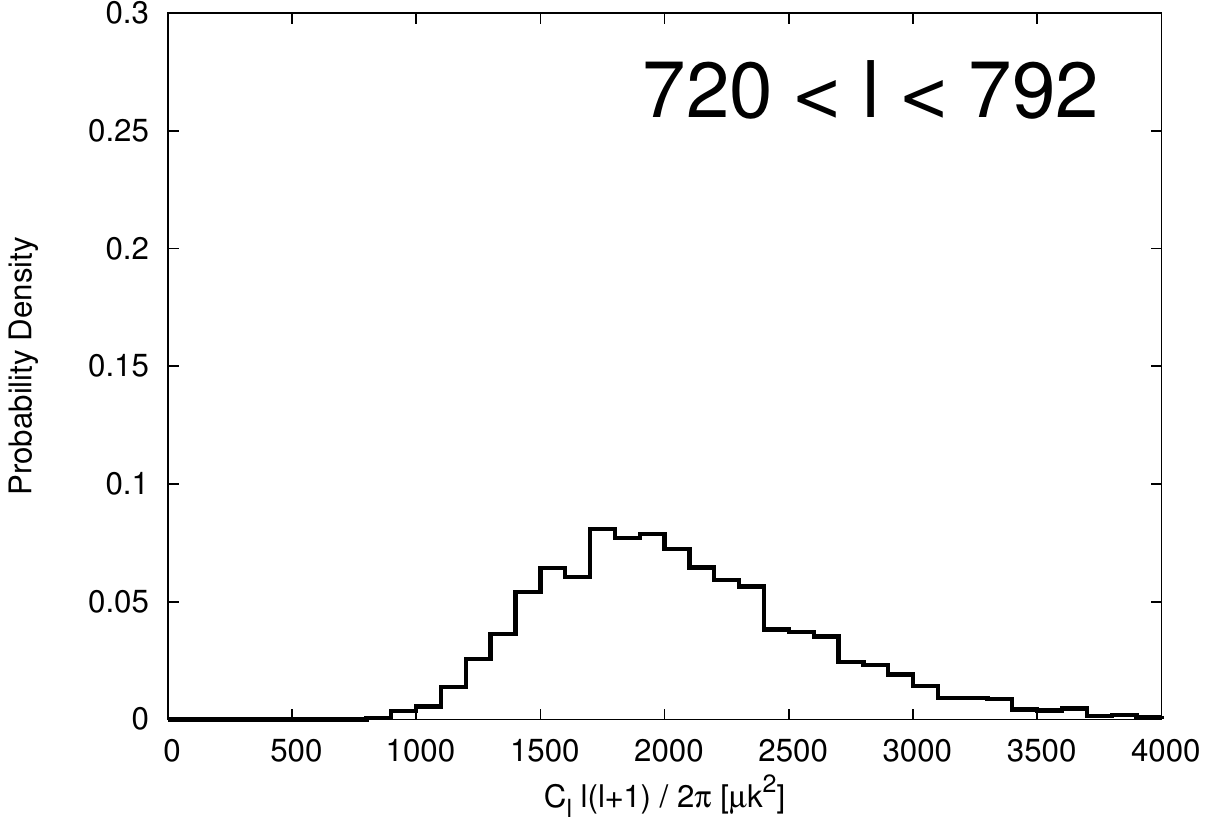}}
   {\includegraphics[width=0.31\textwidth]{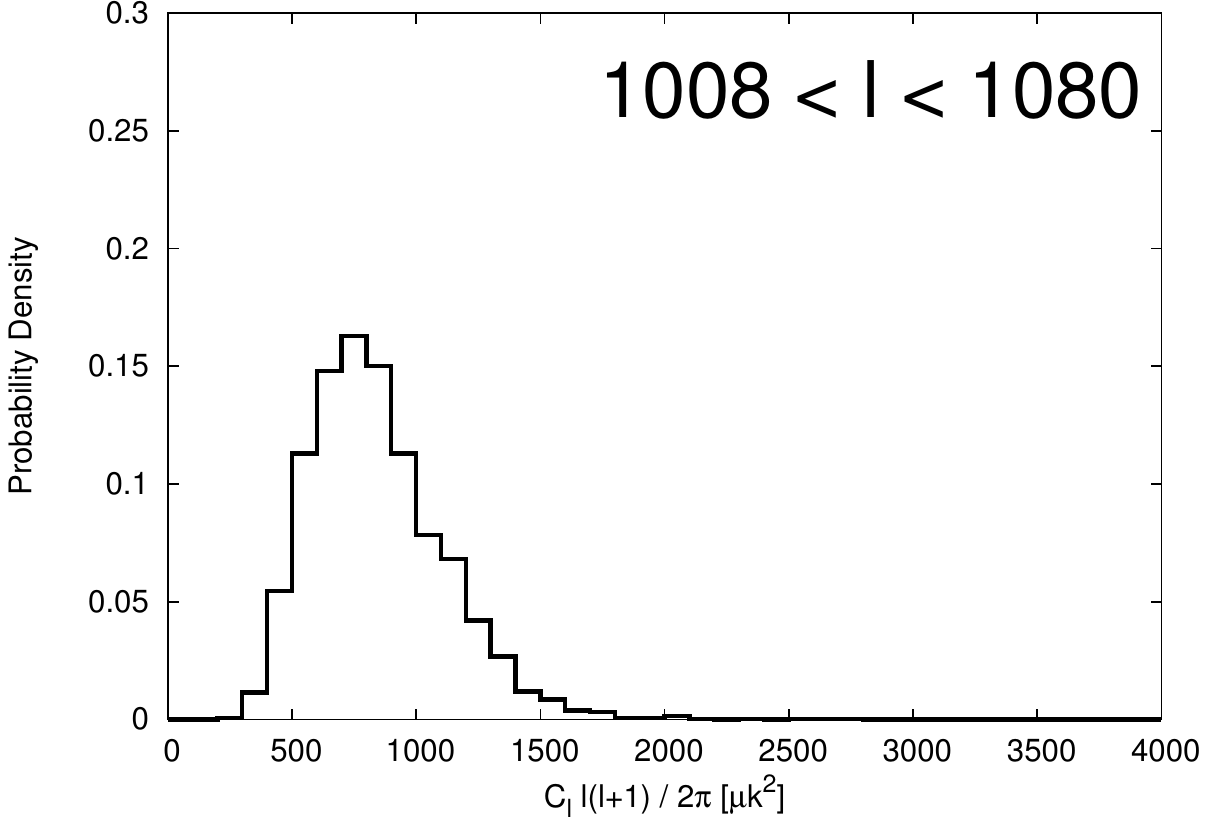}}
   {\includegraphics[width=0.31\textwidth]{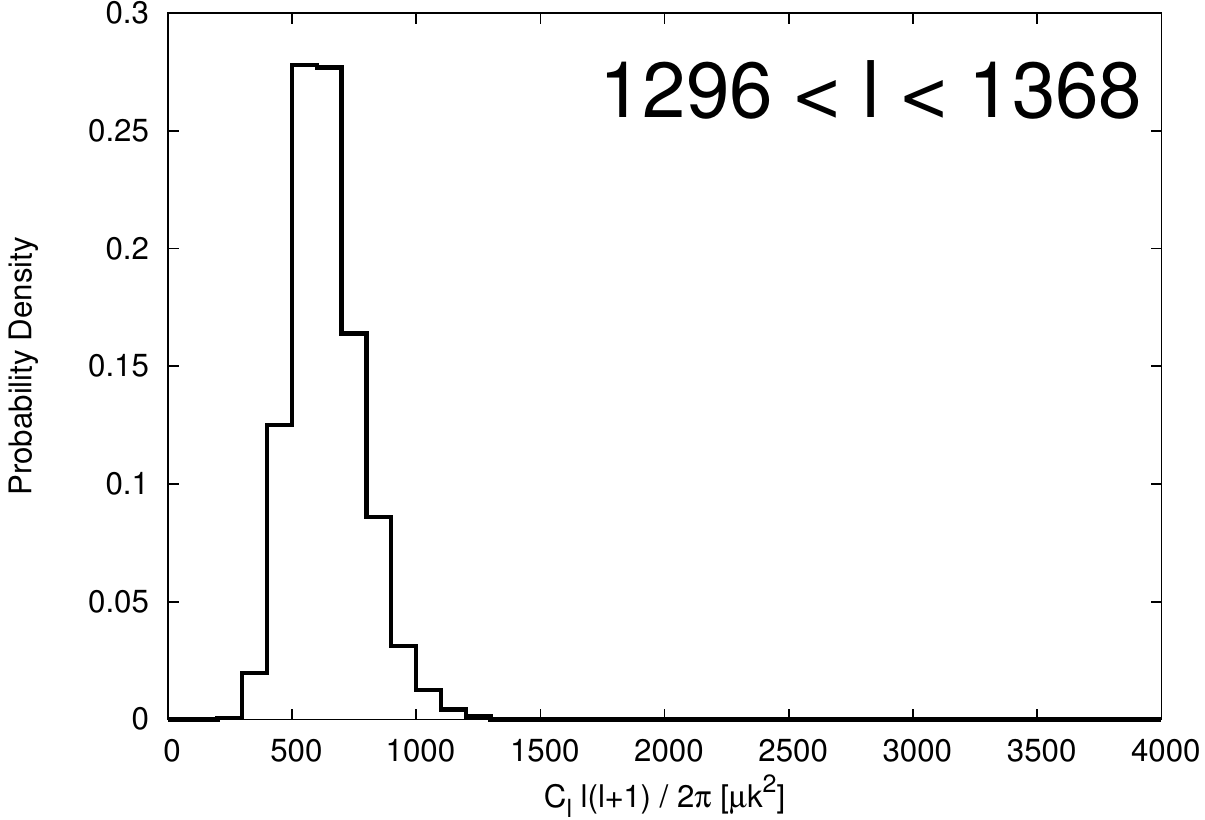}}
  \caption{Histograms of marginalized probability densities for 
            a selection of $\ell$-bins with varying levels of $uv$-plane 
            coverage (see Figure~\ref{fig:skycoverage}). 
            Note that all samples outside of the 
            plotted range are collected in the edge bins.} 
\label{fig:clhist}
\end{figure*}

Figure~\ref{fig:clhist} shows individual marginalized probability 
densities for a selection of $\ell$-bin. We immediately note the 
shape of each probability distribution  matches qualitatively that of an
inverse gamma distribution, as expected. We see the damaging effects of the
lack of $uv$-plane coverage especially in the case of $\ell \sim 300$.  Here,
the lack of any input signal leads to a uniformly decreasing posterior 
probability density function (pdf) with
a very long power-law tail to large power. This can be summarized as an upper
bound on the power in that bin. In other bins, slightly reduced coverage
combined with the effects of noise yields wide and noisy distributions.

\begin{figure} \centering 
  {\includegraphics[width=\columnwidth]{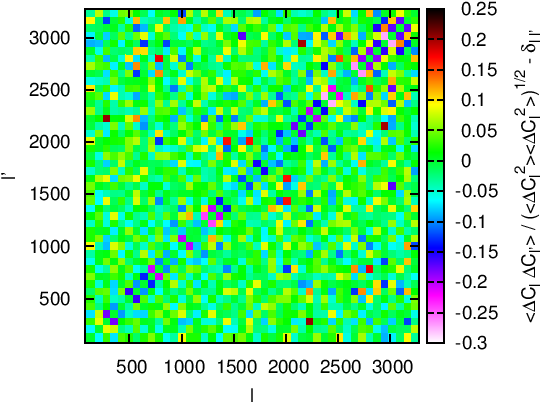}}
  \caption{Power spectrum correlation matrix. Shown are the correlations 
           between all pairs of angular power spectrum bins. 
           Diagonal elements are
           suppressed to enhance the visibility of off-diagonal correlations.
           This matrix highlights the mode coupling due to the beam 
           as correlations between adjacent bins. We also see slight 
           (anti-)correlations of adjacent angular power spectrum bins 
           due to discrete sampling of the angular 
           power spectrum.}
\label{fig:corrmatrix}
\end{figure}

Beyond the marginalized posteriors for each angular power spectrum bin amplitude, the
posterior samples also contain higher-order information.
Figure~\ref{fig:corrmatrix} shows the two-point correlations between all pairs
of angular power spectrum bins. For this plot, we suppress the diagonal components so
that we can more easily examine the cross-bin correlations.  While the
correlation matrix is somewhat noisy, we immediately see the effects of the
reduced sky coverage due to a finite beam width as correlations between 
adjacent modes. This beam leads to reduced
Fourier space resolution. The correlation 
data are informative about regions of the signal data $d$ which
are larger than a single angular power spectrum bin. 
This slight oversampling of the
power leads to a tendency of bins to be anti-correlated with each other.
Anti-correlation occurs because data constrain power in regions of the signal 
and if a particular bin scatters high, the inference in adjacent 
bins reduces in power to remain consistent with the 
data~\citep{Elsner2012}. 

\subsection{Signal Reconstruction}

\begin{figure*} 
\centering  
\subfigure[Signal Sample, Iteration 0]{
    \includegraphics[width=0.45\textwidth]{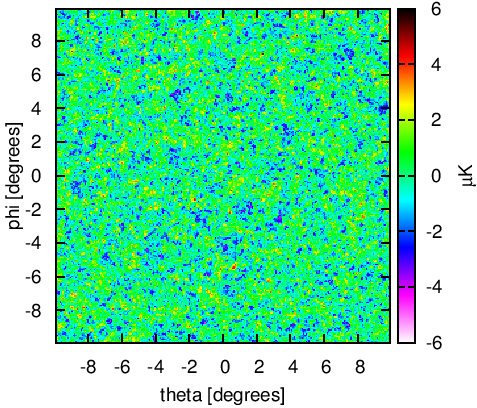}
  }  
  \subfigure[Wiener-Filtered Signal, Iteration 0]{
    \includegraphics[width=0.45\textwidth]{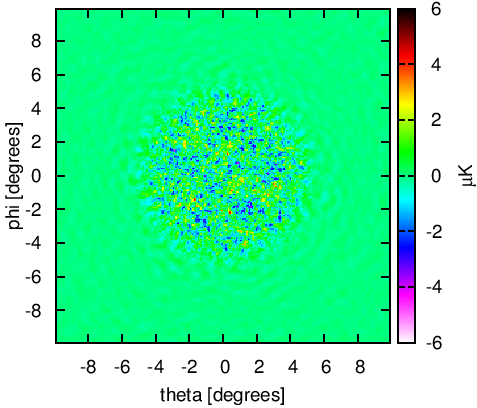}
  }   
  \subfigure[Signal Sample, Iteration 1000]{
    \includegraphics[width=0.45\textwidth]{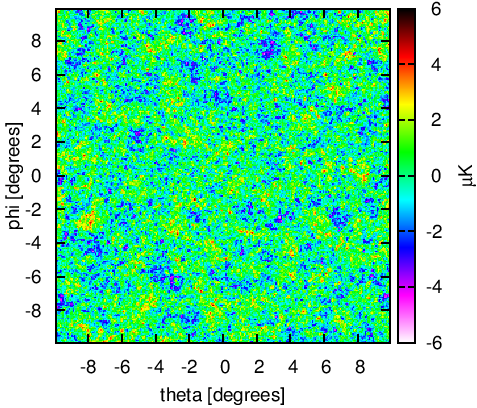}
  }  
  \subfigure[Wiener-Filtered Signal, Iteration 1000]{
    \includegraphics[width=0.45\textwidth]{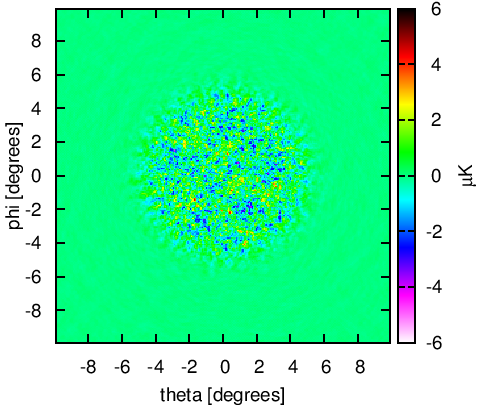}
  }    
  
  \caption{Posterior signal samples
           at various stages of the Gibbs sampling algorithm. 
           Shown are
           (a) signal sample with fluctuations at step 0 
              after the burn-in phase,
           (b) Wiener-filtered map at step 0 after the burn-in phase,      
           (c) signal sample with fluctuations at step 1000, and
           (d) Wiener-filtered map at step 1000.
          }
\label{fig:skies}
\end{figure*}

Figure~\ref{fig:skies} shows several sample posterior maps (i.e., the solution
$s^{i+1}$ to Equation~\ref{eq:sky}).  We show the signal samples at steps 0 and 1000
after the ``burn-in'' phase. The Gibbs sampler fills in the map outside of the
primary beam with fluctuations consistent with the actual data. While maps made
from interferometric data are often accompanied by an ``effective beam'', each
signal sample presented here has no smoothing. The signal samples are a
population of possible pure signal skies which are consistent with the data.

We can also easily compute Wiener-filtered maps: setting the fluctuation 
vectors $\xi_1=\xi_2=0$ in Equation~(\ref{eq:sky}) and solving 
for $s^{i+1}$ provides the definition of a Wiener-filtered 
map~\citep{Wandelt2004}.
We show these maps for iterations 0 and 1000 also in Figure~\ref{fig:skies}.
The Wiener filter adaptively smooths the map depending on the data support. In
this case, the filter eliminates the fluctuations outside of the primary beam,
where there is no data, and successfully recovers the signal in the observed
region.

\begin{figure}   
\centering
\subfigure[Dirty Map]{
   \includegraphics[width=\columnwidth]{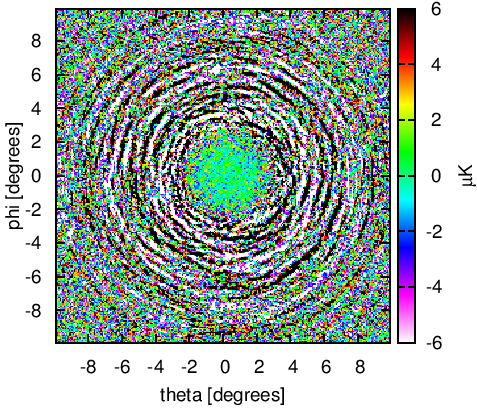}
} 
\subfigure[Final Mean Posterior Map]{
   \includegraphics[width=\columnwidth]{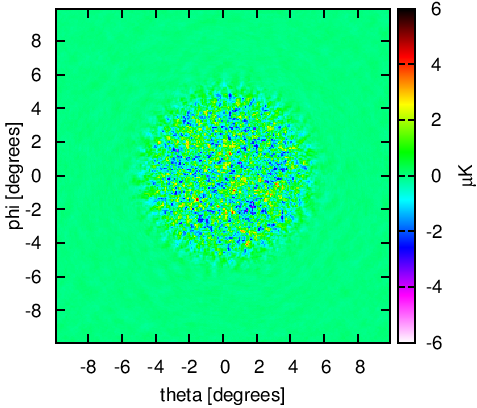}
} 

  \caption{Final map reconstructions. Shown are the ``dirty'' map $F^{-1}
           (I A F~s+n)$ divided by the beam response (a). 
           and the final mean posterior signal map (b).
           We have limited the 
           beam to 1\% of its maximum value for clarity.
           }
\label{fig:finalskies} 
\end{figure}

After sufficient iterations the artificially created fluctuations average out
and we are left with a reliable reconstruction of the input signal within the
area of the primary beam, which we show in Figure~\ref{fig:finalskies}. We
accompany this final mean posterior signal map with the ``dirty'' map, which is
$F^{-1} (I A F~s+I n)$. Note that the dirty map contains artifacts such as
grating rings and sidelobes which are absent from our posterior map.
Also, the dirty map completely misses the strongest fluctuations. 

\begin{figure}   
  \includegraphics[width=\columnwidth]{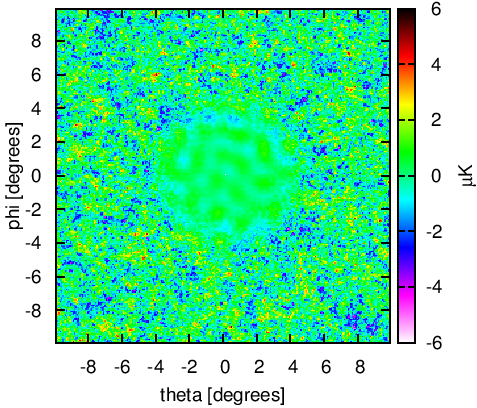}
  \caption{Final map residual: shown is the difference between the 
           final mean posterior signal map 
           and the input map.}
\label{fig:residualmap} 
\end{figure}

We show the difference between the final mean posterior signal map to the input 
map in Figure~\ref{fig:residualmap}. We see that medium-scale fluctuations 
remain; however, the amplitude of these fluctuations 
are below the typical scale of fluctuations in the input signal and 
are consistent with the uncertainties in the median-$\ell$ reconstructed 
power spectrum (Figure~\ref{fig:powerspec}), which are caused by 
incomplete $uv$-plane coverage.
Outside the beam our final posterior map is nearly zero (as desired), 
leading to a reconstruction of the input map in the residual.

\section{Conclusion}
\label{sec:conclusions}

We have successfully demonstrated the technique of Gibbs sampling as applied to
interferometric observations. Our approach accounts for realistic
interferometer features, such as baseline-dependent noise, mode coupling 
due to a finite beam size,
and incomplete coverage of the $uv$-plane. We have presented an example of CMB
angular power spectrum estimation and signal reconstruction of a moderately large ($n_p
= 256^2$) mock data set directly applicable to current and near-future
missions.

A complete interferometric analysis tool for a realistic future mission must include several features
not studied here: multiple frequencies, polarization~\citep{Chiueh2002},
mosaicking~\citep{Bunn2007a}, a true spherical sky~\citep{BunnEmoryF.2008}, and
wide bandwidths~\citep{Subrahmanyan2004}. Building a complete pipeline is beyond the scope of the present study which focused on the ability of Gibbs sampling to explore the full shape of the joint, multivariate, non-Gaussian  posterior pdf for the angular power spectrum and the signal map given the data.

Within the Bayesian framework, strictly speaking, angular power spectrum estimation
only makes sense if Gaussianity and statistical isotropy is assumed.  An
isotropic Gaussian process prior is a highly accurate model of the CMB
anisotropies  and  therefore the detailed shape of the derived angular power spectrum
posterior is of great physical interest for that application. 

While we have focused our example on CMB observations, our basic technique can
be generalized to a wide variety of interferometer observations, such as the
planned study of the 21~cm signal from the epoch of
reionization~\citep{Morales2004}, which can provide useful constraints on
cosmological parameters~\citep{McQuinn2006}.  
In  applications to non-CMB data sets the Gaussian
process prior may not strictly hold, but can be motivated from a maximum
entropy argument: given only a model of the mean and covariance (defined by the
angular power spectrum) of a data set, the least informative completion to a full
probabilistic model  yields a Gaussian process prior. In that case the method
still maps out the angular power spectrum likelihood taking into account all modeled
signals and imperfections in the data at the two-point correlation level. 

In
the derivation of the Gibbs sampling method using the conditional densities of
the posterior distribution, the signal angular power spectrum samples are  paired with
optimal (minimum-variance) signal reconstructions assuming that same power
spectrum.  Averaged along the chain the algorithm therefore simultaneously
discovers the correlation structure in the data, reconstructs optimal sky maps,
and explores the range of uncertainty left by having finite
amounts of imperfect data.
The mean posterior signal
reconstruction can be viewed as a non-linear generalization of the least-squares
optimal signal reconstruction provided by the Wiener filter without requiring
an \emph{a priori} choice of the signal covariance.
 
In any case, the assumption of statistical isotropy in the covariance model
still holds in a standard FRW cosmology even for applications to 21 cm data on
any given redshift slice.  To tackle such a difficult analysis, our Gibbs
sampling approach must be extended to deal with the significant galactic 
and extragalactic
foregrounds present~\citep{Santos2005,Furlanetto2006,Bernardi2009,Liu2012} 
and to
include in the solution the full three spatial dimensions of the data 
and frequency
dependence of the angular power spectrum.

By extending the Gibbs sampling framework to interferometric observations, we
have demonstrated the validity of this method in this regime and its power in
providing computationally efficient $\mathcal{O}(n_p\log n_p)$ angular power spectrum
analysis and signal reconstruction. Our method is able to fully explore 
the angular power spectrum likelihood shape in an optimal fashion. 
However, even with this scaling extremely
large data sets can still take weeks to analyze with a single machine.
Fortunately there are many opportunities for parallelism. For example,
independent chains can be run as separate execution threads in a straightforward
manner. Also, we developed our code with the open-source PETSc library
~\citep{petsc-efficient,petsc-user-ref,petsc-web-page} and the MPI-parallelized
version of FFTW~\citep{Frigo2005}, meaning that maps can be divided up among
multiple processors with communication handled by the internal libraries. This
Bayesian framework can therefore plausibly cope with the extreme data analysis
requirements of future CMB or 21 cm missions.

\section*{Acknowledgments}

The authors acknowledge support from NSF Grant AST-0908902 and useful
discussions with our collaboration partners Ted Bunn, Ata Karakci, Andrei
Korotkov, Peter Timbie, Greg Tucker, and Le Zhang. Computing resources were provided by the
University of Richmond under NSF Grant 0922748. Our implementation of the Gibbs
sampling algorithm uses the open-source PETSc library and FFTW.

\bibliography{InterGibber}		
\bibliographystyle{apj}	
\nocite{*}

\end{document}